\documentclass[twocolumn,english]{revtex4}
\usepackage[T1]{fontenc}
\usepackage[latin9]{inputenc}
\usepackage{color}
\usepackage{babel}
\usepackage{booktabs}
\usepackage{bm}
\usepackage{multirow}
\usepackage{amstext}
\usepackage[unicode=true]
 {hyperref}
\usepackage{graphicx}

\makeatletter

\@ifundefined{textcolor}{}
{%
 \definecolor{BLACK}{gray}{0}
 \definecolor{WHITE}{gray}{1}
 \definecolor{RED}{rgb}{1,0,0}
 \definecolor{GREEN}{rgb}{0,1,0}
 \definecolor{BLUE}{rgb}{0,0,1}
 \definecolor{CYAN}{cmyk}{1,0,0,0}
 \definecolor{MAGENTA}{cmyk}{0,1,0,0}
 \definecolor{YELLOW}{cmyk}{0,0,1,0}
}

\makeatother

\begin{document}
\title{Resolution of the identity approximation applied to PNOF correlation
calculations}
\author{Juan Felipe Huan Lew-Yee$^{1}$, Mario Piris$^{2,3,*}$, Jorge M. del
Campo$^{1,}$\bigskip{}
}

\email{mario.piris@ehu.eus, jmdelc@unam.mx}

\address{$^{1}$Departamento de Física y Química Teórica, Facultad de Química,
Universidad Nacional Autónoma de México, Mexico City, C.P. 04510,
México\\
$^{2}$Kimika Fakultatea, Euskal Herriko Unibertsitatea (UPV/EHU)
and Donostia International Physics Center (DIPC), 20018 Donostia,
Euskadi, Spain. \\
$^{3}$Basque Foundation for Science (IKERBASQUE), 48013 Bilbao, Euskadi,
Spain.\bigskip{}
}
\begin{abstract}
In this work, the required algebra to employ the resolution of the
identity approximation within Piris Natural Orbital Functional (PNOF)
is developed, leading to an implementation named DoNOF-RI. The arithmetic
scaling is reduced from fifth-order to fourth-order, and the memory
scaling is reduced from fourth-order to third-order, allowing significant
computational time savings. After the DoNOF-RI calculation has fully
converged, a restart with four-center electron repulsion integrals
can be performed to remove the effect of the auxiliary basis set incompleteness,
quickly converging to the exact result. The proposed approach has
been tested on cycloalkanes and other molecules of general interest
to study the numerical results as well as the speed-ups achieved by
PNOF7-RI when compared with PNOF7.\bigskip{}

 \bigskip{}
\end{abstract}
\maketitle

\section{Introduction}

Recently \cite{Piris2021}, an open-source implementation of natural
orbital functional (NOF) based methods has been made available to
the scientific community. The associated computer program \textcolor{cyan}{\href{http://github.com/DoNOF/DoNOFsw}{\underline{DoNOF}}}
is designed to solve the energy minimization problem of an approximate
NOF which describes the ground-state of an N-electron system in terms
of the natural orbitals (NOs) and their occupation numbers (ONs).
Approximate NOFs have demonstrated \cite{Mitxelena2019} to be more
accurate than density functionals for highly multi-configurational
systems, and scale better with the number of basis functions than
correlated wave-function methods. A detailed account of the state
of the art of the NOF-based methods can be found elsewhere \cite{Goedecker2000,Piris2007,Piris2014,Pernal2016,Piris2018a}.

A route \cite{PNOF-2006} for the construction of an approximate NOF
involves the employment of necessary N-representability conditions
\cite{Mazziotti2012} for the two-particle reduced density matrix
(2RDM) reconstructed in terms of the one-particle reduced density
matrix (1RDM). Appropriate 2RDM reconstructions have led to different
implementations known in the literature as PNOFi (i=1-7) \cite{PNOF1-2005,PNOF2-2007,PNOF3-2010,PNOF4-2010,PNOF5-2011,Piris2013,PNOF6-2014,PNOF7-2017}.
This family of functionals provide an efficient way of including dynamic
and static correlation with chemical accuracy in many cases \cite{Lopez2010,Piris2016}.
It has recently been shown \cite{Mitxelena2020,Mitxelena2020a} that
PNOF7 is an efficient method for strongly correlated electrons in
one and two dimensions. In addition, the use of perturbative corrections
allow to improve the dynamic correlation in order to achieve a complete
method to describe electron correlated systems \cite{Piris2018,Piris2019}.

In the current implementation, \textcolor{cyan}{\href{http://github.com/DoNOF/DoNOFsw}{\underline{DoNOF}}}
computer code needs to transform the atomic orbital (AO) electron
repulsion integrals (AO-ERIs) into molecular orbital (MO) electron
repulsion integrals (MO-ERIs) in order to evaluate the Coulomb and
exchange integrals required in PNOF. The optimization process involves
searching for ONs, which requires the computation of Coulomb and exchange
matrices in MO representation, and for NOs, which requires computing
Coulomb and exchange matrices in AO representation for each MO. These
procedures have overall fifth-order arithmetic scaling factor. While
this scaling factor is lower compared to other procedures such as
those based on configuration interaction and coupled cluster approaches,
there is still room for improvement.

Resolution of the identity (RI), also known as density fitting \cite{Whitten1973,Dunlap1979,Feyereisen1993},
approximates the product of basis functions as a linear combination
of an auxiliary basis set \cite{Vahtras1993}. It usually reduces the
arithmetic and memory scaling factors, and produces intermediate easy-to-handle
arrays, as has been reported in other methodologies \cite{Calaminici2017,Kendall1997,Sodt2006,Sodt2008,Weigend2000,Sherrill2013,Bozkaya2014,Buu2019,Scheffler2019,Visscher2020,LewYee2020,Nain2020}
such as RI-MP2 \cite{Weigend1997,Weigend1998,Bozkaya2014,Ishikawa2012,Katouda2016,Vogt2008},
DF-MP2 \cite{Bozkaya2014a,Bozkaya2014a}, DF-MP2.5 \cite{Bozkaya2016,Bozkaya2018},
DF-MP3 \cite{Bozkaya2016,Bozkaya2018}, DF-LCCD \cite{Bozkaya2016DF-LCCD},DF-CCSD \cite{Sherrill2013,Bozkaya16a,Bozkaya2020},
and DF-CCSD(T) \cite{Peng2019,Bozkaya2020}. In particular, the use
of the RI approximation in v2RDM-CASSCF calculations \cite{Fosso2016,Mullinax2019}
has been shown, leading to energy expressions and handling of the
MO-ERIs in the optimization procedure different from those necessary
in the PNOF family of functionals. Applying the RI approximation in
PNOF correlation calculations allows faster calculations, decreasing
the arithmetic scale factor of the integral transformation of AO-ERIs
to MO-ERIs from fifth order to fourth order, as shown in this work.

The text is structured as follows. In the second section, the elemental
theory of PNOF formulation is shown and the use of the RI approximation
in the ONs and NOs optimization process is analyzed. In the third
section, the details about the implementation are given. In the fourth
section, the time savings due to the use of the RI approximation as
well as the energy results in standard cycloalkanes test set up to
nine carbon atoms are presented, another relevant molecules such as
oxazole, borazine, coumarin, cyanuryc chloride, benzene, thiepine,
and thieno{[}2,3-b{]}thiophene are also presented. Finally, conclusions
are given in the fifth section.

\section{\label{sec:Theory}Theory}

The ground-state electronic energy of an approximate NOF is given
by the expression 
\begin{equation}
E=2\sum\limits _{p}n_{p}H_{pp}+\sum\limits _{pqrs}D[n_{p},n_{q},n_{r},n_{s}](pq|rs)\label{ENOF}
\end{equation}
where $H_{pp}$ denotes the one-electron matrix elements of the kinetic
energy and outer potential operators, $(pq|rs)$ are the MO-ERIs in
chemists' notation, and $D[n_{p},n_{q},n_{r},n_{s}]$ represents the
reconstructed 2RDM from the ONs $\left\{ n_{p}\right\} $. Restrictions
on the ONs to the range $0\leq n_{p}\leq1$ represent the necessary
and sufficient conditions for ensemble N-representability of the 1RDM
under the normalization condition, $2\sum_{p}n_{p}=\mathrm{N}$.

It is worth noting that any explicit dependence of $D$ on the NOs
$\left\{ \phi_{p}\right\} $ themselves is neglected. Accordingly,
NOs are the MOs that diagonalize the 1RDM of an approximate ground-state
energy, so it is more appropriate to speak of a NOF rather than a
functional of 1RDM due to the explicit dependence on the 2RDM \cite{Piris2018b}.

It is clear that the construction of an N-representable functional
given by Eq.~(\ref{ENOF}) is related to the N-representability problem
of $D$. Using its ensemble N-representability conditions to generate
a reconstruction functional leads to PNOF \cite{PNOF-2006}. This
particular reconstruction is based on the introduction of two auxiliary
matrices $\bm{\Delta}$ and $\bm{\Pi}$ expressed in terms of the
ONs to reconstruct the cumulant part of the 2RDM \cite{Mazziotti1998}.
For the sake of simplicity, let us address only singlet states in
this work. The generalization of our results to spin-multiplet states
\cite{Piris2019} is straightforward. Consequently, energy expression
of Eq.~(\ref{ENOF}) becomes 
\begin{equation}
\begin{array}{c}
E=2\sum\limits _{p}n_{p}H_{pp}+\sum\limits _{qp}\Pi_{qp}L_{pq}\qquad\qquad\\
+\sum\limits _{qp}\left(n_{q}n_{p}-\Delta_{qp}\right)\left(2J_{pq}-K_{pq}\right)
\end{array}\label{PNOF}
\end{equation}
where $J_{pq}$, $K_{pq}$, and $L_{pq}$ are Coulomb, exchange, and
exchange-time-inversion integrals \cite{Piris1999}. Note that $L_{pq}=K_{pq}$
for real MOs as developed in this work. Therefore, only two-index
$J_{pq}$ and $K_{pq}$ integrals are necessary due to our approximation
for the 2RDM. Appropriate forms of matrices $\bm{\Delta}$ and $\bm{\Pi}$
lead to different implementations known as PNOFi (i=1-7). Remarkable
is the case of PNOF5 which turned out to be strictly pure N-representable
\cite{Piris2013e}.

In the current implementation, minimization of the energy $E\left[\left\{ n_{p}\right\} ,\left\{ \phi_{p}\right\} \right]$
is performed under orthonormality requirement for real NOs, whereas
ONs conform to the ensemble N-representability conditions. The solution
is established by optimizing the functional of Eq.~(\ref{PNOF})
with respect to the ONs and to the NOs, separately \cite{Piris2009}.

In \textcolor{cyan}{\href{http://github.com/DoNOF/DoNOFsw}{\underline{DoNOF}}} \cite{Piris2021},
the Coulomb integrals are built according to the equation 
\begin{eqnarray}
J_{pq} & = & \sum_{\mu\nu}P_{\mu\nu}^{p}J_{\mu\nu}^{q}\label{Jpq}\\
 & = & \sum_{\mu}C_{\mu p}\sum_{\nu}C_{\nu p}\sum_{\sigma}C_{\sigma q}\sum_{\lambda}C_{\lambda q}(\mu\nu|\sigma\lambda)\>\>\nonumber 
\end{eqnarray}
where the indices $\mu$, $\nu$, $\sigma$, $\lambda$ label AOs
of dimension $N_{b}$, and $(\mu\nu|\sigma\lambda)$ is an AO-ERI.
Hence, ${\bm{J}}^{q}$ is the Coulomb matrix in AO basis for the MO
$\phi_{q}$, and ${\bm{P}}^{p}$ is computed by means of the MO coefficient
matrix, ${\bm{C}}$, as 
\begin{equation}
P_{\mu\nu}^{p}=C_{\mu p}C_{\nu p}\>\>.\label{Pmat}
\end{equation}
Similarly, the integrals are defined as 
\begin{eqnarray}
K_{pq} & = & \sum_{\mu\sigma}P_{\mu\sigma}^{p}K_{\mu\sigma}^{q}\label{Kpq}\\
 & = & \sum_{\mu}C_{\mu p}\sum_{\sigma}C_{\sigma p}\sum_{\nu}C_{\nu q}\sum_{\lambda}C_{\lambda q}(\mu\nu|\sigma\lambda)\>\>\nonumber 
\end{eqnarray}
where ${\bm{K}}^{q}$ is the exchange matrix in AO basis for the MO
$\phi_{q}$.

From Eqs. (\ref{Jpq}) - (\ref{Kpq}), we observe that the four-index
transformation of the ERIs generally scales as ${N}_{b}^{5}$. In
the occupancy optimization, this operation is carried out once for
fixed orbitals, however, in the orbital optimization it is necessary
to perform this transformation every time orbitals change, which is
a time-consuming process.

\begin{table}[htbp]
\caption{Algorithm used to compute ${\bm J}$ and ${\bm K}$ in the occupancy optimization, and ${\bm J}^q$ and ${\bm K}^q$ in the orbital optimization.\bigskip{}}
\begin{tabular}{c|c|l|ll}
\toprule
\hline
& \multirow{2}{*}{Step} & \multicolumn{1}{c|}{\multirow{2}{*}{Operation}} &  \multicolumn{2}{c}{Scaling}\\ [5pt] \cline{4-5}
&  &  & Memory & Arithmetic \\ [5pt]
\midrule
\multirow{2}{*}{Common} & 0 & Evaluation of $(\mu\nu | \sigma\lambda)$ & $ N_b^4 $ & $N_b^4$ \\
 & 1 & $P_{\mu \nu}^p  = C_{\mu p} C_{\nu p}$ & $ N_b^2 N_{\Omega}$ & $N_b^2 N_{\Omega}$ \\ [5pt]
\midrule
\multirow{2}{*}{$J_{pq}$} & 2 & $J_{\mu \nu}^q = \sum_{\sigma \lambda} P_{\sigma \lambda}^q (\mu \nu | \sigma \lambda)$ & $N_b^2 N_{\Omega}$ & $ N_b^4 N_{\Omega}$ \\[5pt]
 & 3 & $J_{pq} = \sum_{\mu \nu} P_{\mu \nu}^p J^q_{\mu \nu}$ & $N_{\Omega}^2$ & $ N_b^2 N_{\Omega}^2$ \\[5pt]
\midrule
\multirow{2}{*}{$K_{pq}$} & 2 & $K_{\mu \sigma}^q = \sum_{\nu \lambda} P_{\nu \lambda}^q (\mu \nu | \sigma \lambda)$ & $N_b^2 N_{\Omega}$ & $ N_b^4 N_{\Omega}$ \\[5pt]
 & 3 & $K_{pq} = \sum_{\mu \sigma} P_{\mu \sigma}^p K^q_{\mu \sigma}$ & $N_{\Omega}^2$ & $N_b^2 N_{\Omega}^2$ \\[5pt]
\hline
\bottomrule
\end{tabular}
\label{tab:J-K-4c}
\end{table}

It is worth noting that the last members of the PNOF family, namely
PNOF5-PNOF7, use electron-pairing constraints \cite{Piris2018a}.
Until now, only these NOFs can provide the correct number of electrons
in the fragments after a homolytic dissociation \cite{Matxain2011,Piris2016}.
Moreover, the constrained nonlinear programming problem for the ONs
can be transformed into an unconstrained optimization with the corresponding
saving of computational time. In the case of electron-pairing approaches,
we can additionally reduce the number of orbitals in calculations,
and use just orbitals in the pairing scheme, which we will represent
as $N_{\Omega}$ ($N_{\Omega}\leq N_{b}$). From now on we will focus
on the electron-pairing-based PNOFs.

In Table {\ref{tab:J-K-4c}}, we show the conventional algorithm
used to compute the Coulomb (${\bm{J}}$) and exchange (${\bm{K}}$)
integrals in MO representation, and the Coulomb (${\bm{J}}^{q}$)
and exchange (${\bm{K}}^{q}$) matrices in AO representation for each
orbital $\phi_{q}$. In the last columns, the memory and arithmetic
scaling of the steps are reported. We see that the evaluation of the
AO-ERIs $(\mu\nu|\sigma\lambda)$, labeled as step zero, has an arithmetic
scaling of $N_{b}^{4}$. In the current implementation, they are evaluated
and stored at the beginning, consequently, this step does not contribute
significantly to the computational time. However, its storage represents
the highest memory demand with a memory scaling of $N_{b}^{4}$.

The first step corresponds to the evaluation of ${\bm{P}}$ matrix,
as shown in Eq.~(\ref{Pmat}), which has low arithmetic and memory
scaling factors of $N_{b}^{2}N_{\Omega}$. The second step corresponds
to the evaluation of ${\bm{J}}^{q}$ and ${\bm{K}}^{q}$ matrices
for each MO in AO basis. This is the bottleneck of the current implementation
with an arithmetic scaling factor of $N_{b}^{4}N_{\Omega}$ and memory
scaling of $N_{b}^{2}N_{\Omega}$. Finally, in the third step, ${\bm{J}}$
and ${\bm{K}}$ integrals in MO representation are computed with an
arithmetic scaling factor of $N_{b}^{2}N_{\Omega}^{2}$. The memory
scaling of this step is $N_{\Omega}^{2}$, which is not significant
compared to the other steps.

\begin{figure}
\caption{General scheme of the energy optimization. A guess for ONs and NOs
is considered, then an iterative procedure composed of two independent
optimizations, with respect to ONs and NOs respectively, is performed.
For a more detailed description, see the reference [1].\bigskip{}}
\centering{}{\includegraphics[scale=0.3]{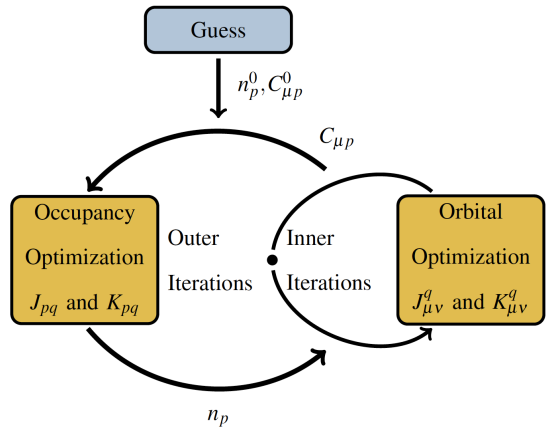}}
\label{fig:outer_inner_iterations}
\end{figure}

As mentioned above, energy minimization is made up of two independent
optimization procedures, an outer one that involves the optimization
of the ONs for fixed orbitals, and an inner one that involves the
optimization of the NOs for fixed occupancies, as shown in Fig. \ref{fig:outer_inner_iterations}.
Both optimizations are iterative procedures in which many inner iterations
are performed per each outer iteration until convergence. In the next
subsections, the introduction of the RI approximation in each optimization
procedure applied to PNOFi (i=5-7) is analyzed. For further reference,
to emphasize the specific functional used, the calculations within
this approach will be labeled as PNOFi-RI (i=5-7), while the global
implementation will be named DoNOF-RI.

\subsection*{Occupancy Optimization with RI}

In \textcolor{cyan}{\href{http://github.com/DoNOF/DoNOFsw}{\underline{DoNOF}}} \cite{Piris2021},
bounds on ${\left\{ n_{p}\right\} }$ are imposed automatically by
expressing the ONs through new auxiliary variables ${\left\{ \gamma_{p}\right\} }$.
In this way, the constrained minimization problem with respect to
ONs for a fixed set of NOs is transformed into an unconstrained minimization
problem with respect to auxiliary $\gamma$-variables.

Since the orbitals do not change, $\bm{J}$ and $\bm{K}$ can be computed
once and stored along the occupancy optimization process of an outer
iteration. The RI approximation can be used to reduce the arithmetic
scaling factors of ${\bm{J}}$ and ${\bm{K}}$ integrals. In this
approximation, the four-center AO-ERI, $(\mu\nu|\sigma\lambda)$,
is expressed using three-center ERIs, $(\mu\nu|k)$, and two-center
ERIs, $(k|l)$, according to the equation 
\begin{equation}
(\mu\nu|\sigma\lambda)=\sum_{k}(\mu\nu|k)\sum_{l}{\bm{G}}_{kl}^{-1}(l|\sigma\lambda)\>\>,\label{AOERI}
\end{equation}
where $k$, $l$ represent functions of the auxiliary basis of dimension
$N_{aux}$, and $\bm{G}$ is a metric matrix defined as $G_{kl}=(k|l)$.
In a symmetric approach, $\mathbf{G}^{-1/2}$ would be computed through
eigenvalue decomposition or singular value decomposition, and multiplied
by the three-center AO-ERIs, however, the metric matrix may be numerically
ill conditioned \cite{LewYee2020}, having small or even negative eigenvalues.
Although this problem might be surpassed truncating eigenvalues below
a certain tolerance, the overall process is slow and may affect the
numerical stability. Recently, a modified Cholesky decomposition has
been applied to factorize the metric matrix and correct the numerical
problems if required \cite{LewYee2020,NainDelesma2020}. In this approach,
the metric matrix is expressed as \cite{laug} 
\begin{equation}
{\bm{G}}=\mathbf{P}\mathbf{L}\mathbf{D}\mathbf{L}^{T}\mathbf{P}^{T}\>\>,
\end{equation}
where $\mathbf{P}$ is a permutation matrix, $\mathbf{L}$ is a lower
triangular matrix, and $\textbf{D}$ is a block diagonal matrix with
blocks of dimension $1\times1$ and $2\times2$ \cite{Bunch1977}.
The eigenvalue spectrum of the $\mathbf{D}$ matrix is analyzed block
by block to correct negative and very small eigenvalues, giving a
corrected matrix, $\mathbf{\tilde{D}}$ \cite{ChengHigham1998}. In
PNOF correlation calculations a symmetric approach results convenient,
thus the $\mathbf{G}$ matrix is expressed as 
\begin{equation}
\mathbf{G}=\mathbf{P}\mathbf{L}{\tilde{\mathbf{D}}}^{1/2}{\tilde{\mathbf{D}}}^{1/2}\mathbf{L}^{T}\mathbf{P}^{T}\>\>,
\end{equation}
the process of decomposing the $\mathbf{D}$ matrix in its eigenvectors
and eigenvalues is fast due to the small dimension of its blocks.
Once the eigenvalues have been corrected, its square root can be evaluated
directly. Then, a $\mathbf{b}$ tensor is found by solving the following
linear equation system 
\begin{equation}
\mathbf{P}\mathbf{L}{\tilde{\mathbf{D}}}^{1/2}\mathbf{b}^{T}=(\mu\nu|k)\>\>.
\end{equation}

Using RI, the Coulomb and exchange integrals can be expressed as 
\begin{eqnarray}
J_{pq} & = & \sum_{l}b_{pp}^{l}b_{qq}^{l}\>\>,\\
K_{pq} & = & \sum_{l}b_{pq}^{l}b_{pq}^{l}\>\>,
\end{eqnarray}
where the change of indices in $\bm{b}$ denotes contractions from
AOs ($\mu$, $\nu$) to MOs ($p$, $q$) according to 
\begin{eqnarray}
b_{p\nu}^{l} & = & \sum_{\mu}C_{\mu p}b_{\mu\nu}^{l}\>\>,\\
b_{pq}^{l} & = & \sum_{\nu}C_{\nu q}b_{p\nu}^{l}\>\>.\label{bpnu}
\end{eqnarray}

An equivalent $\mathbf{b}$ tensor is employed in RI implementations
that use $\mathbf{G}^{-1/2}$, particularly, the equations are similar
to those used in RI-MP2 \cite{Weigend1997,Weigend1998,Bozkaya2014,Ishikawa2012,Katouda2016,Vogt2008}
to build other MO-ERIs.

\begin{table}[htbp]
\caption{Algorithm used to compute $\bm{J}$ and $\bm{K}$ in the occupancy optimization with RI. Formal memory scaling is shown. However, to optimize memory usage, the contraction of $\bm{b}$ tensor for $\bm{J}$ and $\bm{K}$ (steps 2, 3, and 4) are carried out simultaneously for each $l$, such that the dimension of the auxiliary basis does not affect the memory scaling. \bigskip{}}
\begin{tabular}{c|c|l|ll}
\toprule
\hline
& \multirow{2}{*}{Step} & \multicolumn{1}{c|}{\multirow{2}{*}{Operation}} &  \multicolumn{2}{c}{Scaling}\\ [5pt] \cline{4-5}
&  &  & Memory & Arithmetic \\ [5pt]
\midrule
\multirow{4}{*}{Common} & 0 & Evaluation of $(\mu\nu | k)$ & $ N_b^2 N_{aux} $ & $N_b^2 N_{aux}$ \\[5pt]
 & 1 & Solve $\mathbf{P} \mathbf{L} {\tilde {\mathbf{D}}}^{1/2} \mathbf{b}^T $ & $ N_b^2 N_{aux}$ & $N_b^2N_{aux}^2$ \\[5pt]
 & 2 & $b_{p \nu}^l = \sum_{\mu} C_{\mu p} b_{\mu \nu}^l$ & $N_b N_{aux} N_{\Omega}$ & $N_b^2 N_{aux} N_{\Omega}$ \\[5pt]
 & 3 & $ b_{pq}^l = \sum_\nu C_{\nu q} b_{p \nu}^l$ & $ N_{aux} N_{\Omega}^2$ & $ N_b N_{aux} N_{\Omega}^2$ \\[5pt] 
\midrule
$J_{pq}$ & 4 & $J_{pq} = \sum_{l} b_{pp}^l b_{qq}^l$ & $N_{\Omega}^2$ & $ N_{aux} N_{\Omega}^2$ \\[5pt]
\midrule
$K_{pq}$ & 4 & $K_{pq} = \sum_{l} b_{pq}^l b_{pq}^l$ & $ N_{\Omega}^2$ & $ N_{aux} N_{\Omega}^2$ \\[5pt]
\hline
\bottomrule
\end{tabular}
\label{tab:J-K-3c-occ}
\end{table}

The memory and arithmetic scaling factors of the Eqs. (\ref{AOERI})-(\ref{bpnu})
with the RI approximation are shown in Table~\ref{tab:J-K-3c-occ}.
The zero step corresponds to the evaluation of the $(\mu\nu|k)$ AO-ERIs,
and the first step corresponds to solve the linear equation system
for the $\bm{b}$ tensor with a memory scaling factor of $N_{b}^{2}N_{aux}$
and arithmetic scaling factor of $N_{b}^{2}N_{aux}^{2}$. Assuming
that enough memory is available to store the $\bm{b}$ tensor in AO
representation, this step can be performed only once at the beginning
of the calculation; hence, although the first step has the largest
memory scaling, it does not pose a problem through the iterative process.
The second step is the contraction of an index of the $\mathbf{b}$
tensor from AO to MO with memory scaling of $N_{b}N_{aux}N_{\Omega}$
and arithmetic scaling of $N_{b}^{2}N_{aux}N_{\Omega}$, being the
most demanding step per outer iteration; in the third step the remaining
atomic orbital is contracted with arithmetic scaling of $N_{b}N_{aux}N_{\Omega}^{2}$
and memory scaling of $N_{aux}N_{\Omega}^{2}$ respectively. Finally,
in step four, the $\mathbf{b}$ tensor is used to build the Coulomb
and exchange integrals with arithmetic scaling of $N_{aux}N_{\Omega}^{2}$
and memory scaling of $N_{\Omega}$. The overall procedure has a fourth-order
arithmetic scaling of $N_{b}^{2}N_{aux}N_{\Omega}$.

\subsection*{Orbital Optimization with RI}

In the inner optimization procedure of the current implementation
(see Fig. \ref{fig:outer_inner_iterations}), the energy minimization
is performed with respect to real MOs under the requirement of orthonormality,
and considering a fixed set of ONs. In general, an approximate NOF
is not invariant with respect to an orthogonal transformation of the
orbitals. Consequently, orbital optimization cannot be reduced to
a pseudo-eigenvalue problem like in the Hartree-Fock approximation.

In \textcolor{cyan}{\href{http://github.com/DoNOF/DoNOFsw}{\underline{DoNOF}}} \cite{Piris2021},
the optimal NOs are obtained by iterative diagonalizations of a symmetric
matrix ${\bm{F}}^{\lambda}$ determined by the Lagrange multipliers
$\left\{ \lambda_{pq}\right\} $ associated to the orthonormality
conditions. A remarkable advantage of this procedure is that the orthonormality
constraints are automatically satisfied. Unfortunately, the diagonal
elements cannot be determined from the symmetry property of $\bm{\lambda}$,
so this procedure does not provide a generalized Fockian in the conventional
sense. Nevertheless, $\left\{ F_{pp}^{\lambda}\right\} $ may be determined
with the help of an aufbau principle \cite{Piris2009}.

Thus, the orbital optimization requires to calculate $\left\{ \lambda_{pq}\right\} $
in each step of the inner iterations in order to determine the symmetric
matrix ${\bm{F}}^{\lambda}$. Since orbitals change in each step,
$\bm{J}^{q}$ and $\bm{K}^{q}$ must be recomputed in each inner iteration.
Many inner iterations are performed per outer iteration, so the computation
of these matrices in the orbital optimization is the most important
contribution to the computational time of the present algorithm.

\begin{table}[htbp]
\caption{Algorithm used to compute $\bm{J}^q$ and $\bm{K}^q$ in the orbital optimization with RI. Formal memory scaling is shown. However, to optimize memory usage, the contraction of $\bm{b}$ tensor for $\bm{J}^q$ (steps 2, 3, and 4) and $\bm{K}^q$ (steps 2 and 3) are carried out simultaneously for each $l$, such that the dimension of the auxiliary basis does not affect the memory scaling. \bigskip{} }
\begin{tabular}{c|c|l|ll}
\toprule
\hline
& \multirow{2}{*}{Step} & \multicolumn{1}{c|}{\multirow{2}{*}{Operation}} &  \multicolumn{2}{c}{Scaling}\\ [5pt] \cline{4-5}
&  &  & Memory & Arithmetic \\ [5pt]
\midrule
 & 0 & Evaluation of $(\mu\nu | k)$ & $ N_b^2 N_{aux} $ & $N_b^2 N_{aux}$ \\ [5pt]
Common & 1 & Solve $\mathbf{P} \mathbf{L} {\tilde {\mathbf{D}}}^{1/2} \mathbf{b}^T $ & $ N_b^2 N_{aux}$ & $N_b^2N_{aux}^2$ \\ [5pt]
 & 2 & $b_{q\nu}^l = \sum_{\mu} C_{\mu q} b_{\mu \nu}^l$ & $N_b N_{aux}N_{\Omega}$ & $ N_b^2 N_{aux} N_{\Omega}$ \\[5pt] \midrule
\multirow{3}{*}{$J_{\mu\nu}^q$} & 3 & $b_{qq}^l = \sum_{\nu} C_{\nu q} b_{q\nu}^l$ & $N_{aux}N_{\Omega}$ & $ N_bN_{aux}N_{\Omega}$ \\[5pt]
 & 4 & $J_{\mu \nu}^q = \sum_{l} b_{qq}^l b_{\mu\nu}^l$ & $N_b^2 N_{\Omega}$ & $ N_b^2 N_{aux} N_{\Omega}$ \\[5pt]
\midrule
$K_{\mu\nu}^q$ & 3 & $K_{\mu \nu}^q = \sum_l b_{q\mu}^l b_{q\nu}^l$ & $N_b^2 N_{\Omega}$ & $ N_b^2 N_{aux} N_{\Omega}$ \\[5pt]
\hline
\bottomrule
\end{tabular}
\label{tab:J-K-3c-orb}
\end{table}

The RI approximation can also be applied in this case, using the procedure
shown in Table~\ref{tab:J-K-3c-orb}. The zero and first steps evaluate
the $(\mu\nu|k)$ AO-ERIs and the $\bm{b}$ tensor in AO basis, both
are common steps shared with the occupancy optimization and performed
at the beginning of the calculation. In the second step, an index
of the $\mathbf{b}$ tensor is contracted from AO to MO with arithmetic
scaling of $N_{b}^{2}N_{aux}N_{\Omega}$. In the third step of the
Coulomb procedure, an additional contraction is performed for the
$\mathbf{b}$ tensor. Finally, in the last steps of both the Coulomb
and exchange procedures, the intermediate tensors are multiplied to
compute $\bm{J}^{q}$ and $\bm{K}^{q}$. The algorithm reduces the
arithmetic scaling factor of orbital optimization to the fourth-order
($N_{b}^{2}N_{aux}N_{\Omega}$), as in the previous case. Hence, an
overall reduction of the arithmetic scaling factor from fifth-order
to the fourth-order, and of the memory scaling factor from fourth-order
to the third-order is achieved due to the RI approximation.

\begin{table}[htbp]
\caption{Comparison of the energies (Hartrees) obtained with PNOF7, PNOF7-RI using aug-cc-pVDZ/GEN-A2* for the cycloalkanes test. Mean diff: $2.2\times10^{-4}$ \bigskip{}}
\begin{tabular}{ll|rr}
    \toprule
    \hline
     \multicolumn{2}{c}{Molecule} &  \multicolumn{1}{c}{$E_{PNOF7}$} & \multicolumn{1}{c}{$\Delta E_{PNOF7-RI}$\footnote{Positive differences mean that PNOF7-RI energy is above than the PNOF7 energy.}} \\
    \hline
    Cyclopropane & (C$_{3}$H$_{6}$) & -117.228991 & $1.5\times 10^{-4}$ \\
    Cyclobutane & (C$_{4}$H$_{8}$) & -156.328758 & $1.9\times 10^{-4}$ \\
    Cyclopentane & (C$_{5}$H$_{10}$) & -195.449913 & $2.5\times 10^{-4}$ \\
    Cyclohexane & (C$_{6}$H$_{12}$) & -234.549938 & $2.2\times 10^{-4}$ \\
    Cycloheptane & (C$_{7}$H$_{14}$) & -273.630436 & $2.3\times 10^{-4}$ \\
    Cyclooctane & (C$_{8}$H$_{16}$) & -312.714209 & $2.4\times 10^{-4}$ \\
    Cyclononane & (C$_{9}$H$_{18}$) & -351.799073 & $2.9\times 10^{-4}$ \\
    \hline
    \bottomrule
\end{tabular}
\label{tab:energy}
\end{table}

\section{\label{sec:Computational_Details}Computational Details}

The proposed PNOFi-RI (i=5-7) algorithm was implemented in a modified
version of the DoNOF software \cite{Piris2021} using Cartesian Gaussian
basis functions and MPI parallelization, leading to a new implementation
labeled as DoNOF-RI.

We assume that there is enough memory available to compute at the
beginning all the required AO-ERIs as well as the $\mathbf{b}$ tensor
on the atomic basis, and store them for use along the calculation.
Operations of optimization procedures correspond only to arithmetic
manipulations and not to AO-ERI evaluations. Four-center AO-ERIs,
$(\mu\nu|\sigma\lambda)$, have been screened to discard those lower
than $10^{-9}$. This approach has been taken to reduce the arithmetic
scaling when four center ERIs are used \cite{Almlof1982,Haser1989,Whitten1973,Maurer2012}.
All results shown in this article were calculated using 24 threads
of an Intel Xeon Gold 5118 CPU. Basis sets were taken from the basis
set exchange \cite{Schuchardt2007,Feller1996,Pritchard2019} www.basissetexchange.org
website.

\section{\label{sec:Results}Results}

\begin{figure}[htbp] 
\caption{Analysis of occupancy (top panel) and orbital optimizations (bottom panel) for PNOF7 and PNOF7-RI computing time using aug-cc-pVDZ/GEN-A2{*}. Achieved speed-up is presented over each pair of bars. \bigskip{}}
\label{fig:SpeedUps}
\centering{}{\includegraphics[scale=0.5]{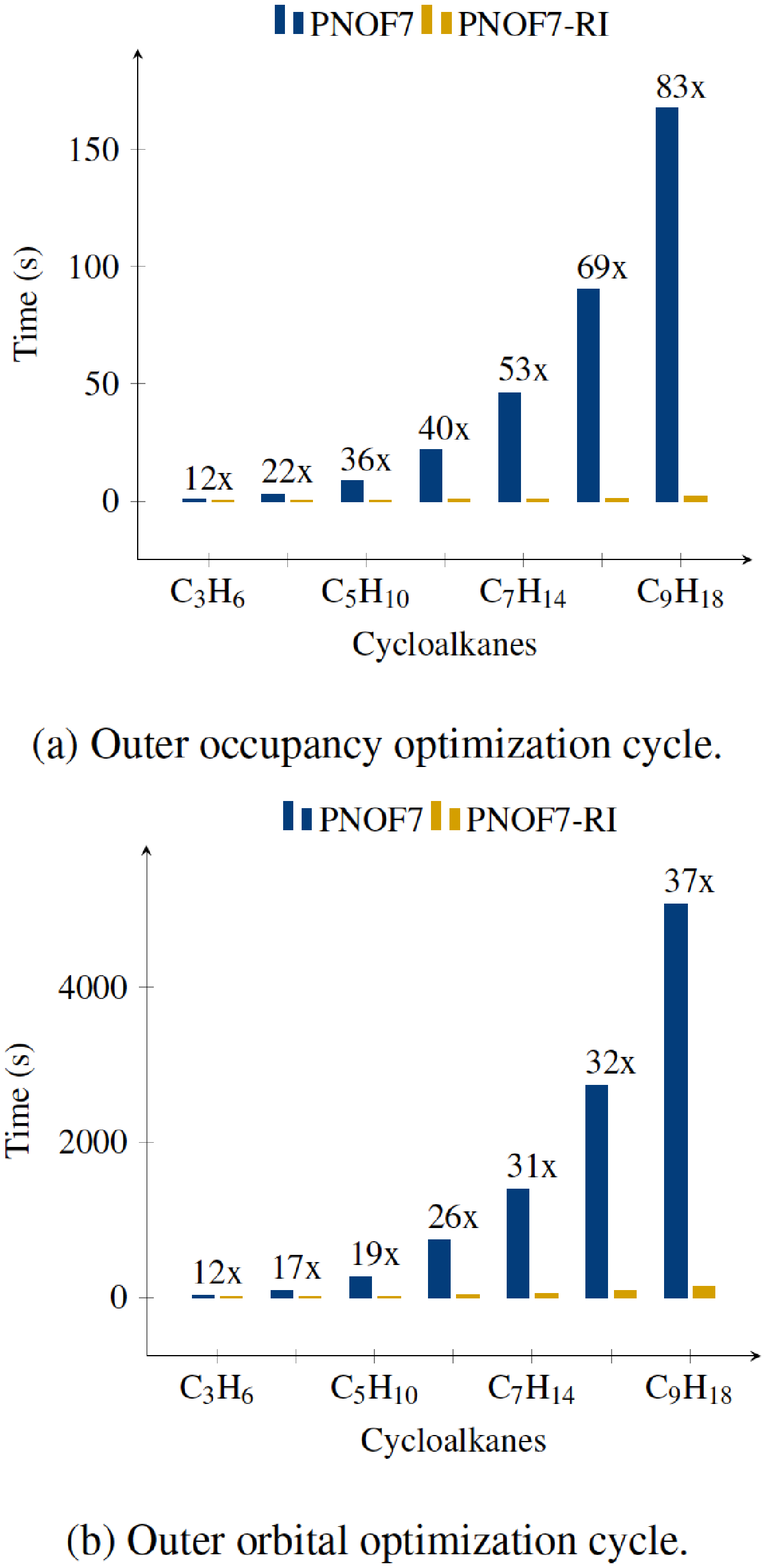}}
\end{figure}

Single point energy calculations were performed to study the numerical
stability and speed-up achieved with the DoNOF-RI implementation.
The structures were optimized with Psi4 software \cite{Parrish2017}
using M06-2X \cite{Zhao2008} and aug-cc-pVDZ/aug-cc-pVDZ-jkfit \cite{Weigend2002}
basis set. Initial auxiliary variables ${\left\{ \gamma_{p}^{0}\right\} }$
corresponding to a Fermi--Dirac distribution of ${\left\{ n_{p}^{0}\right\} }$
were employed. For NOs, the guess MOs were taken from a Hartree-Fock
calculation.

\begin{table*}
    \centering
    \caption{Comparison of the energies (Hartrees) obtained with PNOF7, PNOF7-RI using cc-pVTZ/GEN-A2* for molecules of general interest.  Mean diff: $3.1\times10^{-3}$ \bigskip{}}    
    \begin{tabular}{ll|rrr}
    \toprule
    \hline
         \multicolumn{2}{c}{Molecule} & \multicolumn{1}{c}{$E_{PNOF7}$} & \multicolumn{1}{c}{$\Delta E_{PNOF7-RI}$}\footnote{Positive differences mean that the PNOF7-RI energy is above than the PNOF7 energy.} & Speed-up\footnote{Global speed-up per outer iteration} \\
    \hline
        Oxazole & (C$_3$H$_3$NO) & -244.980370 & 8.8$\times 10^{-4}$ & 23 \\
        Borazine & (B$_3$H$_3$N$_3$) & -241.487944 & 7.0$\times 10^{-4}$ & 19 \\
        Coumarin & (C$_9$H$_6$O$_2$) & -494.724761 & 1.7$\times 10^{-3}$ & 19 \\
        Cyanuric Chloride & (C$_3$Cl$_3$N$_3$) & -1655.966373 & 8.0$\times 10^{-3}$ & 23 \\
        Benzene & (C$_6$H$_6$)& -231.058747 & 6.7$\times 10^{-4}$ & 28 \\
        Thiepine & (C$_6$H$_6$S) & -628.585882 & 2.4$\times 10^{-3}$ & 37 \\
        Thieno[2,3-b]thiophene & (C$_6$H$_4$S$_2$) & -1239.953451 & 7.1$\times 10^{-3}$ & 27 \\        
    \hline
    \bottomrule
    \end{tabular}
    \label{tab:pnof7-pnof7ri-general}
\end{table*}

Figure \ref{fig:SpeedUps} presents the computational times of an
outer iteration for occupancy optimization (top panel) as well as
for orbital optimization (bottom panel) from cyclopropane to cyclononane
employing aug-cc-pVDZ basis set \cite{Kendall1992,Dunning1989} and
GEN-A2{*} auxiliary basis set \cite{gen1,gen2,gen3}, which generates
auxiliary basis functions according to the basis set. In both plots,
blue bars represent the elapsed time obtained with PNOF7 and yellow
bars correspond to computed time with PNOF7-RI, the speed-up achieved
by PNOF7-RI with respect to PNOF7 is presented over each pair of bars.
The different sizes of the blue bars compared to the yellow bars makes
evident the different arithmetic scaling factors between PNOF7 and
PNOF7-RI. For the smallest cycloalkane tested, C$_{3}$H$_{6}$, an
outer iteration of PNOF7-RI is 12 times faster than the equivalent
iteration in PNOF7, in the other hand, for the largest cycloalkane
tested, C$_{9}$H$_{18}$, PNOF7-RI is 83 and 37 times faster for
occupancy and orbital optimization respectively. Speed-ups for occupancy
and orbital optimization behave accordingly to the described arithmetic
scaling factors, since the final steps of the integral evaluation
for the orbital optimization shown in Table~\ref{tab:J-K-3c-orb}
have slightly higher arithmetic scaling factors than the final steps
of the integral evaluation in the occupancy optimization described
in Table~\ref{tab:J-K-3c-occ}.

Although a significant reduction of computational time has been achieved,
it is important to analyze the numerical impact of the RI approximation
applied to PNOF7 on the final energy values. For this purpose, the
NO's and ON's of the converged PNOF7-RI calculation have been used
to restart the calculation using four center ERIs, namely, a PNOF7
calculation. The results are presented in Table~\ref{tab:energy},
where the PNOF7 energy and PNOF7-RI energy difference for each cycloalkane
is tabulated. It can be seen that PNOF7-RI allows achieving a general
accuracy between three and four decimal places, with a mean difference
of $2.2\times10^{-4}$ Hartrees. In all cases a restart of the PNOF7-RI
calculation converged to the PNOF7 energy in at most two outer iterations,
allowing for a PNOF7 result in a reduced amount of time.

The described restarting procedure using cc-pVTZ/GEN-A2{*} basis sets
for molecules of general interest has been performed. The results
are shown in Table~\ref{tab:pnof7-pnof7ri-general}, where the PNOF7
energy is shown with the corresponding deviation of the PNOF7-RI result.
The minimum error of $6.7\times10^{-4}$ corresponds to the benzene
molecule, and the maximum error of $1.7\times10^{-3}$ corresponds
to the coumarin molecule. The global times of an outer iteration of
PNOF7-RI and PNOF7 were compared and the result can be seen in the
column labeled as speed-up, where it is shown that PNOF7-RI is 37
times faster than PNOF7 for the case of the thiepine, as well as important
speed-ups for the other cases. Overall, the results prove that DoNOF-RI
allows to compute medium size molecules of general interest.

\bigskip{}

\section{\label{sec:Conclusions}Conclusions}

The resolution of the identity approximation has proved to be significant
to decrease the arithmetic and memory scaling factors of the PNOFi
(i=5-7) functionals, leading to the DoNOF-RI implementation. The generality
of the algorithm proposed here makes it applicable to all approximate
natural orbital functionals known so far. While having an acceptable
deviation of the final energy value, the solution for the natural
orbitals and occupation numbers can be used as a start guess for a
regular PNOF calculation with convergence in few iterations. Consequently,
DoNOF-RI provides a way of reaching accurate results in a reduced
amount of time, allowing PNOFi (i=5-7) functionals to be used to study
systems of general interest.

\bigskip{}\bigskip{}

\begin{acknowledgments}
J. F. H. Lew-Yee with CVU number 867718 gratefully thanks CONACyT
for PhD scholarship. J. M. del Campo acknowledges funding from CONACyT
project CB-2016-282791, PAPIIT-IN114418 and computing resources from
LANCAD-UNAM-DGTIC-270 project. M.P. acknowledges the financial support
of MCIU/AEI/FEDER, UE (PGC2018-097529-B-100) and Eusko Jaurlaritza
(Ref. IT1254-19).
\end{acknowledgments}


\end{document}